\documentclass[12pt]{article}
\usepackage{amssymb, latexsym}

\evensidemargin\oddsidemargin                  
\newtheorem{remark}{Remark}
\newtheorem{theorem}{Theorem}
\newtheorem{corollary}{Corollary}
\newtheorem{lemma}{Lemma}

\begin{document}

\bigskip

\begin{center}
{\LARGE States of quantum systems and their liftings}

\bigskip

{\large Joachim Kupsch}\footnote{%
e-mail: kupsch@physik.uni-kl.de}

{\large Fachbereich Physik, Universit\"at Kaiserslautern\\[0pt]
D-67653 Kaiserslautern, Germany}

\medskip

{\large Oleg G. Smolyanov}\footnote{%
e-mail: smolyan@mail.ru}{\large \ and Nadejda A. Sidorova}\footnote{%
e-mail: nadja@sidorova.mccme.rssi.ru}

{\large Faculty of Mechanics and Mathematics,\\[0pt]
Moscow State University, 119899 Moscow, Russia}

\medskip
\end{center}

\begin{abstract}
Let $H_{1}$, $H_{2}$ be complex Hilbert spaces, $H$ be their Hilbert tensor
product and let $\mathrm{{tr}_{2}}$ be the operator of taking the partial trace
of trace class operators in $H$ with respect to the space $H_{2}$. The
operation $\mathrm{{tr}_{2}}$ maps states in $H$ (i.e. positive trace class
operators in $H$ with trace equal to one) into states in $H_{1}$. In this
paper we give the full description of mappings that are linear right inverse
to $\mathrm{{tr}_{2}}$. More precisely, we prove that any affine mapping $%
F(W)$ of the convex set of states in $H_{1}$ into the states in $H$ that is
right inverse to $\mathrm{{tr}_{2}}$ is given by $W\mapsto W\otimes D$ for
some state $D$ in $H_{2}$.

In addition we investigate a representation of the quantum mechanical state
space by probability measures on the set of pure states and a representation 
-- used in the theory of stochastic Schr\"{o}dinger equations -- by probability
measures on the Hilbert space. We prove that there are no affine mappings from the state
space of quantum mechanics into these spaces of probability measures. 
\end{abstract}

                      
\section{\textbf{INTRODUCTION}}

In quantum mechanics the states of a physical system are given by the
statistical operators or density matrices in the Hilbert space associated to
this system. The state of a subsystem is uniquely calculated as the reduced
statistical operator by the partial trace. But it seems that the inverse
problem: to define a linear mapping from the set of states of a subsystem to
the set of states of an enlarged system such that the reduced state
coincides with the original state, has not been studied systematically in
the literature. In this article we want to investigate this lifting problem
of states and the adjoint problem of reducing observables in some detail.

In the sequel all Hilbert spaces are assumed to be complex (and separable).
For any Hilbert space we denote by $\mathcal{L}(H)$ the (complex) vector
space of all linear bounded operators in $H$; by $\mathcal{L}^{a}(H)$ we
denote the real vector subspace of $\mathcal{L}(H)$ consisting of all
self-adjoint operators from $\mathcal{L}(H)$, by $\mathcal{L}^{+}(H)$ we
denote the cone of positive operators within $\mathcal{L}(H)$ (and hence
within $\mathcal{L}^{a}(H)$). The (complex) vector space of all trace class
operators in $H$ is denoted by $\mathcal{L}_{1}(H)$. In addition we use the
following notations: $\mathcal{L}_{1}^{+}(H)=\mathcal{L}^{+}(H)\cap \mathcal{%
L}_{1}(H)$, $\mathcal{L}_{1}^{a}(H)=\mathcal{L}_{1}^{+}(H)\cap \mathcal{L}%
^{a}(H)$, and $\mathcal{D}(H)$ is the convex set of all operators from $%
\mathcal{L}_{1}^{+}(H)$ having trace equal to one. If $H$ is the Hilbert
space associated to a physical system, then the elements of $\mathcal{L}%
^{a}(H)$ represent the (bounded) observables of the system, the elements of $%
\mathcal{D}(H)$ represent (mixed and pure) states, and the closed subset $%
\mathcal{P}(H)\subset \mathcal{D}(H)$ of rank one projection operators
represents the pure states.

If $\mathcal{S}$ and $\mathcal{E}$ are physical systems with Hilbert spaces $%
H_{S}$ and $H_{E}$, then the Hilbert space of the composite system --
denoted by $\mathcal{S}\times \mathcal{E}$ -- of these systems is the
Hilbert tensor product of Hilbert spaces $H_{S}$ and $H_{E}$, i.e. $%
H=H_{S}\otimes H_{E}$. The scalar product in $H$ is written as $\langle \ ,\
\rangle _{H}$; the corresponding notations are used for scalar products in $%
H_{S}$ and $H_{E}$. Hence $\mathcal{S}$ is a subsystem of the quantum system 
$\mathcal{S}\times \mathcal{E}$, and the system $\mathcal{E}$ can be
interpreted as an environment of $H_{S}$. For any state $W\in \mathcal{D}(H)$
of the total system $\mathcal{S}\times \mathcal{E}$ the state of the system $%
\mathcal{S}$ -- called the reduced state -- is given by the partial trace $%
\mathrm{tr}_{H_{E}}W$ $\in \mathcal{D}(H_{S}).$ This partial trace is
uniquely defined for all $W\in \mathcal{L}_{1}(H)$ as the operator $\mathrm{%
tr}_{H_{E}}W\in \mathcal{L}_{1}(H_{S})$ which satisfies the identity $%
\langle \mathrm{tr}_{H_{E}}Wx_{1},x_{2}\rangle _{H_{S}}=\sum_{j}\langle
W(x_{1}\otimes e_{j}^{E}),(x_{2}\otimes e_{j}^{E})\rangle _{H}$ for an
orthonormal basis $\left\{ e_{j}^{E}\right\} $ of $H_{E}$ and all $%
x_{1},x_{2}\in H_{S}$. The mapping $W\rightarrow \mathrm{tr}_{H_{E}}W,\;%
\mathcal{L}_{1}(H)\rightarrow \mathcal{L}_{1}(H_{S})$, is obviously linear
and continuous.

By the partial trace we can calculate the Schr\"{o}dinger dynamics of the
subsystem $\mathcal{S}$ -- the so called reduced dynamics -- from the
Schr\"{o}dinger dynamics of the whole system $\mathcal{S}\times \mathcal{E}$%
. But in general, this dynamics does not depend linearly on the initial
state of the subsystem, see Remark \ref{dynamics}. In order to obtain the
linear dependence one has to find a linear solution for the lifting problem,
which can be formulated as follows. For any state $W_{S}\in \mathcal{L}%
_{1}(H_{S})$ to find a state $F(W_{S})\in \mathcal{L}_{1}(H)$ such that $%
\mathrm{tr}_{H_{E}}F(W_{S})=W_{S}$; such a mapping $F_{S}$ is called the
lifting.

The simplest solution of this problem is given by the mapping $F_{D}:%
\mathcal{L}_{1}(H_{S})\longrightarrow \mathcal{L}_{1}(H)$, $W\mapsto
W\otimes D$, where $D$ is an element of $\mathcal{L}_{1}(H_{E})$, which is
usually called a reference state. This choice is well known from the theory
of open systems, see e.g. \cite{Favre/Martin:1968},\cite{Martin:1979},\cite
{Davies:1976}.

The main theorem of the paper -- Theorem 1 of the next section -- implies
that actually any linear lifting coincides with $F_{D}$, for some $D$.

\begin{remark}
\label{Gleason}The vector space $\mathcal{L}^{a}(H)$ of bounded observables
can be identified with the space of continuous affine linear functionals on
the state space $\mathcal{D}(H)$ equipped with the topology induced by the
trace norm $\left\| _{.}\right\| _{1}$ of $\mathcal{L}_{1}(H)\supset 
\mathcal{D}(H)$, see e.g. \cite{Reed/Simon:1972}. Affine linearity means
that such a functional $f:\mathcal{D}(H)\longrightarrow \mathbb{R}$ respects
the mixing property: $f(\alpha W_{1}+\beta W_{2})=\alpha f(W_{1})+\beta
f(W_{2})$ for $0\leq \alpha ,\beta \leq 1$ with $\alpha +\beta =1$, and $%
W_{1},W_{2}\in \mathcal{D}(H)$. In fact, any such a functional can be
uniquely extended to a continuous $\mathbb{C}$-linear functional $\bar{f}:%
\mathcal{L}_{1}(H)\longrightarrow \mathbb{C}$, see e.g. \cite{Guz:1974}. Since $%
\mathcal{L}(H)$ is dual to $\mathcal{L}_{1}(H)$, with the duality pairing 
\begin{equation}
\mathcal{L}(H)\times \mathcal{L}_{1}(H)\longrightarrow \mathbb{C}:(A,W)\mapsto
\langle A,W\rangle \equiv \mathrm{tr}_{H}AW,  \label{dual}
\end{equation}
there exists $A_{f}\in \mathcal{L}(H)$ such that for any $W\in \mathcal{L}%
_{1}(H)$ the identity $\overline{f}(W)=\mathrm{tr}_{H}WA_{f}$ is true.

On the other side, according to Gleason's theorem \cite{Gleason:1957}, the
state space $\mathcal{D}(H)$ can be identified with the set of linear
functionals $\omega :\mathcal{L}(H)\longrightarrow \mathbb{C}$ having the
following properties:

(1) if $A\in \mathcal{L}^{+}(H)$ then $\omega (A)\geq 0$;

(2) $\omega (Id)=1;$

(3) $\omega (\sum_{j}P_{j})=\sum_{j}\omega (P_{j})$ for any finite or
countable family of mutually orthogonal projectors.

For any $\omega $ which satisfies these constraints there exists an element $%
W_{\omega }\in \mathcal{D}(H)$ such that $\omega (A)=\mathrm{tr}%
_{H}W_{\omega }A$ is true for all $A\in \mathcal{L}(H)$. The natural norm of
the state space is $\sup_{\left\| A\right\| =1}\left| \omega (A)\right| $
which coincides with the trace norm of $W_{\omega }$.
\end{remark}

\begin{remark}
\label{dynamics}The time evolution of a composite system with Hilbert space $%
H=H_{S}\otimes H_{E}$ in the Schr\"{o}dinger picture is given by a family $%
\Phi _{t},\,t\in \mathbb{R}$ or $t\in \mathbb{R}_{+}$, of continuous affine linear
mappings $\Phi _{t}:\mathcal{D}(H)\longrightarrow \mathcal{D}(H)$. We
normalize these evolutions by $\Phi _{0}(W)=W$. The affine linear mappings $%
\Phi _{t}$ can be extended to $\mathbb{C}$-linear mappings on $\mathcal{L}%
_{1}(H)$, again denoted by $\Phi _{t}$. In the usual case of a Hamiltonian
(unitary) dynamics we have $\Phi _{t}(W)=U(t)WU^{+}(t)$ with the unitary
group $U(t)$ on $H$ generated by the Hamiltonian. But more general
evolutions like semigroups are admitted in the sequel. The mappings $\Phi
_{t}$ have unique extensions to continuous $\mathbb{C}$-linear mappings $%
\overline{\Phi }_{t}$ of $\mathcal{L}_{1}(H)$ into $\mathcal{L}_{1}(H)$. The
duality (\ref{dual}) then allows to determine the Heisenberg evolution, a
family $\Psi _{t}$ of continuous linear operators on $\mathcal{L}(H)$. Any
Schr\"{o}dinger evolution $\Phi _{t}$ on $\mathcal{D}(H_{S}\otimes H_{E})$
induces a unique time evolution $\rho _{t}=tr_{H_{E}}\Phi _{t}(W)$ of the
system $H_{S}$. In order to obtain a linear dependence on the initial state $%
\rho =\rho _{t=0}$ we need an affine linear mapping $F$ of $\mathcal{D}%
(H_{S})$ into $\mathcal{D}(H_{S}\otimes H_{E})$. Then the mapping $\rho
\mapsto W=F(\rho )\mapsto \rho _{t}=tr_{H_{E}}\Phi _{t}(W)$ is a linear time
evolution on $\mathcal{D}(H_{S})$. This time evolution has the correct
initial condition $\rho _{t=0}=\rho $ if $F$ satisfies the constraint $%
\mathrm{tr}_{H_{E}}F(\rho )=\rho $. The Heisenberg dynamics of the system
then follows from the duality (\ref{dual}) applied to $\mathcal{L}(H_{S})$
and $\mathcal{L}_{1}(H_{S})$.
\end{remark}

The paper is organized a follows. In Sec. 2 we prove the main result of the
paper -- Theorem 1 -- describing all linear liftings. In Sec. 3 we consider
a theorem -- Theorem 2 -- that is in a sense dual to Theorem 1 and describes
a reduction of observables of the system $H$ to observables of the system $%
H_{S}$.

In the final Sec. 4 we consider the case of a classical state space, i. e. a
space of probability measures, and the representation of the quantum
mechanical state space $\mathcal{D}(H)$ by probability measures either on
the set of pure states -- the Choquet representation -- or on the Hilbert
space -- a representation used in the theory of stochastic Schr\"{o}dinger
equations. The space $\mathcal{D}(H)$ is a convex set with the closed set $%
\mathcal{P}(H)$ of pure states as extremal points. Any $W\in \mathcal{D}(H)$
can be represented by an integral over the pure states $W=\int_{\mathcal{P}%
(H)}\mu (dP)\,P$, where $\mu (dP)$ is a probability measure on $\mathcal{P}%
(H)$. Since this representation has been derived by Choquet for general
convex sets, see e.g. \cite{Choquet:1969}, we denote the (non-unique)
measure $\mu (dP)$ as Choquet measure of $W$. In Theorem \ref{simplex} we
prove that there does not exist a linear mapping $\gamma $ from the space $%
\mathcal{D}(H)$ into the set of probability measures on the set $\mathcal{P}%
(H)$ such that the measure $\gamma (W)$ is the Choquet measure of the state $%
W\in \mathcal{D}(H)$. This theorem is in fact a consequence of Theorem 1. In
Sec. 4 we deduce Theorem \ref{simplex} from the structural difference
between the classical and the quantum mechanical state spaces. Both these
spaces are convex sets. But the classical state space is a simplex whereas $%
\mathcal{D}(H)$ not, see e.g. \cite{Bel/Cas:1981}. Finally we investigate
the representation of the state space by probability measures on the Hilbert
space. Also in this case the structural difference between the quantum
mechanical state space and the space of probability measures does not allow
an affine linear mapping from $\mathcal{D}(H)$ into the measure space.

\section{\textbf{LINEAR LIFTINGS}}

The main result of the paper is the following theorem.

\begin{theorem}
\label{lifting}Let $F:\mathcal{D}(H_{S})\longrightarrow \mathcal{D}%
(H_{S}\otimes H_{E})$ be an affine linear mapping such that $\mathrm{tr}%
_{H_{E}}F(\rho )=\rho $ for all $\rho \in \mathcal{D}(H_{S})$. Then there
exists an element $\rho _{E}\in \mathcal{D}(H_{E})$ such that $F(\rho )=\rho
\otimes \rho _{E}$.
\end{theorem}

\noindent \textbf{Proof} \ The mapping $F$ can be extended (uniquely) to the 
$\mathbb{C}$-linear mapping of $\mathcal{L}_{1}(H_{S})$ into $\mathcal{L}%
_{1}(H_{S}\otimes H_{E})$ that we shall denote by the same symbol. This
extension has the following properties: 
\begin{equation}
F(\mathcal{L}_{1}^{+}(H_{S}))\subset \mathcal{L}_{1}^{+}(H_{S}\otimes H_{E}),
\end{equation}
\begin{equation}
F(\mathcal{L}_{1}^{a}(H_{S}))\subset \mathcal{L}_{1}^{a}(H_{S}\otimes H_{E});
\end{equation}
we shall use these properties later.

Let $\{e_{i},i\in \mathbb{N}\}$ (respectively, $\{f_{j},j\in \mathbb{N}\}$) be an
orthonormal basis in $H_{S}$ (respectively, in $H_{E}$). Without loss of
generality we assume $H_{S}$ and $H_{E}$ to be infinite-dimensional. Then $H=%
\mathrm{span}\left\{ e_{i}\otimes f_{j},\,i\in \mathbb{N},j\in \mathbb{N}\right\} $%
. We realize $\mathcal{L}_{1}(H_{S})$ as a vector space of complex valued
functions on $\mathbb{N}^{2}$: $g_{ij}\in \mathbb{C}:i\in \mathbb{N},j\in \mathbb{N}$.
Analogously, we realize $\mathcal{L}_{1}(H)=\mathcal{L}_{1}(H_{S}\otimes
H_{E})$ as a vector space of complex valued functions on $\mathbb{N}^{4}$: $%
F(g)_{ij}^{kl},{i,j,k,l\in \mathbb{N}}$. We say that $b_{ij},i,j\in \mathbb{N}$ is
a $(k,l)$-component of $F\in \mathcal{L}_{1}(H)$ and denote it by $(F)^{kl}$
if $F_{ij}^{kl}=b_{ij}$ for all $i,j\in \mathbb{N}$. We say that $F$ has only
the components of some type if all components of other type are equal to
zero. Let us note that 
\begin{equation}
\mathrm{tr}_{H_{E}}F(g)=g\Leftrightarrow \sum_{i=1}^{\infty
}F(g)_{ii}^{kl}=g_{kl},\ \forall k,l\in N.  \label{tr}
\end{equation}
\bigskip

Consider the following basis $\{g^{kl},k\leq l,g^{kl\ast },k<l\}$ in $%
\mathcal{L}_{1}(H_{S})$: 
\[
g_{ij}^{kl}=\left\{ 
\begin{array}{l}
1\ \mbox{ if }(i,j)\in \{(k,k),(k,l),(l,k),(l,l)\}, \\ 
0\ \mbox{ otherwise }
\end{array}
\right. ,\ \ g_{ij}^{kl\ast }=\left\{ 
\begin{array}{l}
1\ \mbox{ if }(i,j)=(k,k), \\ 
i\ \mbox{ if }(i,j)=(k,l), \\ 
-i\ \mbox{ if }(i,j)=(l,k), \\ 
1\ \mbox{ if }(i,j)=(l,l), \\ 
0\ \mbox{ otherwise. }
\end{array}
\right. 
\]

Firstly, all $g^{kl}$ and $g^{kl*}$ are positive operators, therefore $%
F(g^{kl})$ and $F(g^{kl*})$ are also positive and hence $F(g^{kl})_{ii}^{mm}%
\ge 0$ and $F(g^{kl*})_{ii}^{mm}\ge 0$ for all $i,m\in\mathbb{N}$. Due to (\ref
{tr}) $\sum_{i=1}^{\infty}F(g^{kl})_{ii}^{mm}=g^{kl}_{mm}=0$ for $m\neq
k,m\neq l$ and $\sum_{i=1}^{\infty}F(g^{kl*})_{ii}^{mm}=g^{kl*}_{mm}=0$ for $%
m\neq k,m\neq l$. From this follows that 
\begin{equation}  \label{diag}
F(g^{kl})_{ii}^{mm}=0, m\neq k,m\neq l,i\in\mathbb{N}, \ \ \
F(g^{kl*})_{ii}^{mm}=0, m\neq k,m\neq l,i\in\mathbb{N}.
\end{equation}

Secondly, all $g^{kl}$ and $g^{kl\ast }$ are self-adjoint, therefore $%
F(g^{kl})$ and $F(g^{kl\ast })$ are also self-adjoint and hence 
\begin{equation}
F(g^{kl})_{ji}^{nm}=\overline{F(g^{kl})_{ij}^{mn}}\mbox{ and }F(g^{kl\ast
})_{ji}^{nm}=\overline{F(g^{kl\ast })_{ij}^{mn}}\mbox{ for all }%
i,j,k,l,m,n\in \mathbb{N}.  \label{trans}
\end{equation}
\hfill $\Box $ \smallskip

The further proof is organized as follows. First, we show that $F(g^{kk})$
has only $(k,k)$-component (Step 1), $F(g^{kl}),k<l$ (resp., $F(g^{kl*}),k<l$%
) has only $(k,k)$, $(k,l)$, $(l,k)$, $(l,l)$-components (Step 2).
Furthermore, we prove that non-zero components of $F(g^{kl})$ are equal
(Step 3) and that non-zero components of $F(g^{kl*})$ satisfy $%
(F)^{kk}=-i(F)^{kl}=i(F)^{lk}=(F)^{ll}$ (Step 4). Finally, we denote
elements of the only non-zero component of $F(g^{11})$ by $a_{ij}$ and show
that any non-zero component of $F(g^{kl})$ is equal to $a_{ij}$ (Step 5) and
that the non-zero components of $F(g^{kl*})$ satisfy $%
(F)^{kk}=-i(F)^{kl}=i(F)^{lk}=(F)^{ll}=a_{ij}$ (Step 6), which completes the
proof.

In the proof we shall also use the following (obvious) lemma.

\begin{lemma}
\label{pq} Let $a\ge 0,b\ge 0, c\ge 0$ be real numbers; then 
\[
\left\{(t,p)\in\mathbb{R}^2:(1+t)(1+p)\ge 1,t+1\ge 0\right\}\subset 
\]
\[
\left\{(t,p)\in\mathbb{R}^2:(b+at)(b+cp)\ge b^2,b+at\ge 0\right\} 
\]
\[
\Leftrightarrow a=c\le b. 
\]
\end{lemma}

\noindent \textbf{Proof of Theorem 1} (continued)\newline
\textbf{Step 1.} Consider $g^{kk}\in \mathcal{L}_{1}(H_{S}),k\in \mathbb{N}$
and restrict $F(g^{kk})$ to the space $\langle e_{m}\otimes
f_{i},e_{n}\otimes f_{j}\rangle $, where $(m,n)\neq (k,k)$. In this basis $%
F(g^{kk})$ has the form 
\[
\left( 
\begin{array}{cc}
F(g^{kk})_{ii}^{mm} & F(g^{kk})_{ji}^{nm} \\ 
F(g^{kk})_{ij}^{mn} & F(g^{kk})_{jj}^{nn}
\end{array}
\right) . 
\]
Because either $m\neq k$ or $n\neq k$ we have due to (\ref{diag}) that
either $F(g^{kk})_{ii}^{mm}=0$ or $F(g^{kk})_{jj}^{nn}=0$. $F(g^{kk})$ is
positive, hence $%
F(g^{kk})_{ii}^{mm}F(g^{kk})_{jj}^{nn}-F(g^{kk})_{ji}^{nm}F(g^{kk})_{ij}^{mn}\geq 0 
$. Combining this conditions together with (\ref{trans}) we get 
\[
F(g^{kk})_{ij}^{mn}=0\;\forall i,j,m,n\in \mathbb{N},(m,n)\neq (k,k), 
\]
i.e. $F(g^{kk})$ has only $(k,k)$-component. \bigskip

\noindent \textbf{Step 2.} Consider $g^{kl}\in \mathcal{L}_{1}(H_{S}),k,l\in 
\mathbb{N},k<l$ and restrict $F(g^{kl})$ to the subspace ${\langle e_{m}\otimes
f_{i},e_{n}\otimes f_{j}\rangle }$, where $(m,n)$ does not take values $%
(k,k) $, $(k,l)$, $(l,k)$, $(l,l)$. In this basis $F(g^{kl})$ has the form 
\[
\left( 
\begin{array}{cc}
F(g^{kl})_{ii}^{mm} & F(g^{kl})_{ji}^{nm} \\ 
F(g^{kl})_{ij}^{mn} & F(g^{kl})_{jj}^{nn}
\end{array}
\right) . 
\]
Due to the conditions on $(m,n)$ it follows from (\ref{diag}) that either $%
F(g^{kl})_{ii}^{mm}=0$ or $F(g^{kl})_{jj}^{nn}=0$. $F(g^{kl})$ is positive,
hence $%
F(g^{kl})_{ii}^{mm}F(g^{kl})_{jj}^{nn}-F(g^{kl})_{ji}^{nm}F(g^{kl})_{ij}^{mn}\geq 0 
$. Combining this conditions together with (\ref{trans}) we get 
\[
F(g^{kl})_{ij}^{mn}=0\;\forall i,j,m,n\in \mathbb{N}\mbox{ with }(m,n)\notin
\left\{ (k,k),(k,l),(l,k),(l,l)\right\} , 
\]
i.e. $F(g^{kl})$ has only $(k,k)$, $(k,l)$, $(l,k)$, $(l,l)$-components.

Analogously (substituting $g^{kl*}$ for $g^{kl}$) we prove that $F(g^{kl*})$
has only $(k,k)$, $(k,l)$, $(l,k)$, $(l,l)$-components. \bigskip

\noindent \textbf{Step 3.} First, let us show that the main diagonals of the
non-zero components of $F(g^{kl})$ are equal, i.e. $%
F(g^{kl})_{ii}^{kk}=F(g^{kl})_{ii}^{kl}=F(g^{kl})_{ii}^{lk}=F(g^{kl})_{ii}^{ll}. 
$ Restrict $F(g^{kl})$ to the subspace $\langle e_{k}\otimes
f_{i},e_{l}\otimes f_{i}\rangle $. In this basis $F(g^{kl})$ has the form 
\[
\left( 
\begin{array}{cc}
F(g^{kl})_{ii}^{kk} & F(g^{kl})_{ii}^{lk} \\ 
F(g^{kl})_{ii}^{kl} & F(g^{kl})_{ii}^{ll}
\end{array}
\right) . 
\]
This matrix is positive, hence $%
F(g^{kl})_{ii}^{kk}F(g^{kl})_{ii}^{ll}-F(g^{kl})_{ii}^{lk}F(g^{kl})_{ii}^{kl}\geq 0 
$, i.e.\newline
$|F(g^{kl})_{ii}^{kl}|\leq \sqrt{F(g^{kl})_{ii}^{kk}F(g^{kl})_{ii}^{ll}}$
(note that $F(g^{kl})_{ii}^{kk}$ and $F(g^{kl})_{ii}^{ll}$ are real and
non-negative). Due to (\ref{tr}) $\sum_{i=1}^{\infty
}F(g^{kl})_{ii}^{kk}=\sum_{i=1}^{\infty
}F(g^{kl})_{ii}^{kl}=\sum_{i=1}^{\infty
}F(g^{kl})_{ii}^{lk}=\sum_{i=1}^{\infty }F(g^{kl})_{ii}^{ll}=1$ and hence 
\begin{eqnarray*}
1 &=&\sum_{i=1}^{\infty }\mbox{Re}F(g^{kl})_{ii}^{kl}\leq \sum_{i=1}^{\infty
}|F(g^{kl})_{ii}^{kl}|\leq \sum_{i=1}^{\infty }\sqrt{%
F(g^{kl})_{ii}^{kk}F(g^{kl})_{ii}^{ll}} \\
&\leq &\sum_{i=1}^{\infty }\frac{F(g^{kl})_{ii}^{kk}+F(g^{kl})_{ii}^{ll}}{2}%
=1,
\end{eqnarray*}
and therefore all parts of the inequality must be equal. We have 
\[
\sqrt{F(g^{kl})_{ii}^{kk}F(g^{kl})_{ii}^{ll}}=\frac{%
F(g^{kl})_{ii}^{kk}+F(g^{kl})_{ii}^{ll}}{2}\Longrightarrow
F(g^{kl})_{ii}^{kk}=F(g^{kl})_{ii}^{ll} 
\]
and 
\[
\mbox{Re}F(g^{kl})_{ii}^{kl}=|F(g^{kl})_{ii}^{kl}|=F(g^{kl})_{ii}^{kk}%
\Longrightarrow F(g^{kl})_{ii}^{kl}=F(g^{kl})_{ii}^{lk}=F(g^{kl})_{ii}^{kk}. 
\]
Hence the diagonal elements $%
F(g^{kl})_{ii}^{kk}=F(g^{kl})_{ii}^{kl}=F(g^{kl})_{ii}^{lk}=F(g^{kl})_{ii}^{ll} 
$ are equal.

Secondly, let us show that the corresponding non-diagonal elements of the
non-zero components of $F(g^{kl})$ are equal, i.e. $%
F(g^{kl})_{ij}^{kk}=F(g^{kl})_{ij}^{kl}=F(g^{kl})_{ij}^{lk}=F(g^{kl})_{ij}^{ll}, 
$ where $i\neq j$. Denote $a_{i}$=$F(g^{kl})_{ii}^{kk}$. Restrict $F(g^{kl})$
to the subspace $\langle e_{k}\otimes f_{i},e_{k}\otimes f_{j},e_{l}\otimes
f_{j}\rangle $. In this basis $F(g^{kl})$ has the form 
\[
\left( 
\begin{array}{ccc}
F(g^{kl})_{ii}^{kk} & F(g^{kl})_{ji}^{kk} & F(g^{kl})_{ji}^{lk} \\ 
F(g^{kl})_{ij}^{kk} & F(g^{kl})_{jj}^{kk} & F(g^{kl})_{jj}^{lk} \\ 
F(g^{kl})_{ij}^{kl} & F(g^{kl})_{jj}^{kl} & F(g^{kl})_{jj}^{ll}
\end{array}
\right) =\left( 
\begin{array}{ccc}
a_{i} & \bar{y} & \bar{x} \\ 
y & a_{j} & a_{j} \\ 
x & a_{j} & a_{j}
\end{array}
\right) =A. 
\]
If $a_{j}=0$ then obviously $x=y=0$ as A is positive. If $a_{j}\neq 0$ then 
\[
\mbox{ det}A=-y(\bar{y}a_{j}-\bar{x}a_{j})+x(\bar{y}a_{j}-\bar{x}%
a_{j})=-a_{j}|y-x|^{2}\geq 0\Longrightarrow x=y, 
\]
and we have derived $%
F(g^{kl})_{ij}^{kk}=F(g^{kl})_{ij}^{kl}=F(g^{kl})_{ij}^{lk}=F(g^{kl})_{ij}^{ll} 
$ for all $i,j$.\bigskip

\noindent \textbf{Step 4.} Firstly, let us prove $F(g^{kl\ast
})_{ii}^{kk}=-iF(g^{kl\ast })_{ii}^{kl}=iF(g^{kl\ast
})_{ii}^{lk}=F(g^{kl\ast })_{ii}^{ll},$ i.e. that this condition holds on
the main diagonals of non-zero components of $F(g^{kl\ast })$. Analogously
to the previous step, we get $|F(g^{kl\ast })_{ii}^{kl}|\leq \sqrt{%
F(g^{kl\ast })_{ii}^{kk}F(g^{kl\ast })_{ii}^{ll}}$ (note that $F(g^{kl\ast
})_{ii}^{kk}$ and $F(g^{kl\ast })_{ii}^{ll}$ are real and non-negative). Due
to (\ref{tr}) $\sum_{i=1}^{\infty }F(g^{kl\ast
})_{ii}^{kk}=\sum_{i=1}^{\infty }F(g^{kl\ast
})_{ii}^{ll}=1,\,\sum_{i=1}^{\infty }F(g^{kl\ast })_{ii}^{kl}=i$, $%
\sum_{i=1}^{\infty }F(g^{kl\ast })_{ii}^{lk}=-i$ and hence 
\begin{eqnarray*}
1 &=&\sum_{i=1}^{\infty }\mbox{Im}F(g^{kl\ast })_{ii}^{kl}\leq
\sum_{i=1}^{\infty }|F(g^{kl\ast })_{ii}^{kl}|\leq \sum_{i=1}^{\infty }\sqrt{%
F(g^{kl\ast })_{ii}^{kk}F(g^{kl})_{ii}^{ll}} \\
&\leq &\sum_{i=1}^{\infty }\frac{F(g^{kl\ast })_{ii}^{kk}+F(g^{kl\ast
})_{ii}^{ll}}{2}=1,
\end{eqnarray*}
and therefore all parts of inequality must be equal. Analogously to the
previous step we have 
\[
F(g^{kl\ast })_{ii}^{kk}=F(g^{kl\ast })_{ii}^{ll} 
\]
and 
\[
\mbox{Im}F(g^{kl\ast })_{ii}^{kl}=|F(g^{kl\ast })_{ii}^{kl}|=F(g^{kl\ast
})_{ii}^{kk}\Longrightarrow F(g^{kl\ast })_{ii}^{kl}=F(g^{kl\ast
})_{ii}^{lk}=F(g^{kl\ast })_{ii}^{kk} 
\]
and hence $F(g^{kl\ast })_{ii}^{kk}=-iF(g^{kl\ast })_{ii}^{kl}=iF(g^{kl\ast
})_{ii}^{lk}=F(g^{kl\ast })_{ii}^{ll}$.

Secondly, let us show that this property holds also for corresponding
non-diagonal elements of the non-zero components of $F(g^{kl\ast })$, i.e. $%
F(g^{kl\ast })_{ij}^{kk}=-iF(g^{kl\ast })_{ij}^{kl}=iF(g^{kl\ast
})_{ij}^{lk}=F(g^{kl\ast })_{ij}^{ll}$ if $i\neq j$.

Denote $a_{i}=F(g^{kl\ast })_{ii}^{kk}$. Restrict $F(g^{kl\ast })$ to the
subspace ${\langle e_{k}\otimes f_{i},e_{k}\otimes f_{j},e_{l}\otimes
f_{j}\rangle }$. In this basis $F(g^{kl\ast })$ has the form 
\[
\left( 
\begin{array}{ccc}
F(g^{kl\ast })_{ii}^{kk} & F(g^{kl\ast })_{ji}^{kk} & F(g^{kl\ast
})_{ji}^{lk} \\ 
F(g^{kl\ast })_{ij}^{kk} & F(g^{kl\ast })_{jj}^{kk} & F(g^{kl\ast
})_{jj}^{lk} \\ 
F(g^{kl\ast })_{ij}^{kl} & F(g^{kl\ast })_{jj}^{kl} & F(g^{kl\ast
})_{jj}^{ll}
\end{array}
\right) =\left( 
\begin{array}{ccc}
a_{i} & \bar{y} & \bar{x} \\ 
y & a_{j} & ia_{j} \\ 
x & -ia_{j} & a_{j}
\end{array}
\right) =A. 
\]
If $a_{j}=0$ then obviously $x=iy=0$ as A is positive. If $a_{j}\neq 0$ then 
\[
\mbox{ det}A=-y(\bar{y}a_{j}-i\bar{x}a_{j})+x(-i\bar{y}a_{j}-\bar{x}%
a_{j})=-a_{j}|iy-x|^{2}\geq 0\Longrightarrow x=iy, 
\]
and hence $F(g^{kl\ast })_{ij}^{kk}=-iF(g^{kl\ast })_{ij}^{kl}=iF(g^{kl\ast
})_{ij}^{lk}=F(g^{kl\ast })_{ij}^{ll}$ holds for all $i,j$.\bigskip

\noindent \textbf{Step 5.} Firstly, let us show that the main diagonals of
non-zero components of all $F(g^{kl}),k\leq l,$ are equal, i.e. we have to
prove 
\[
F(g^{kk})_{ii}^{kk}=F(g^{ll})_{ii}^{ll}=F(g^{kl})_{ii}^{kl}\;\mbox{ for all }%
\,k<l. 
\]
Consider $g(t)=g^{kl}+tg^{kk}+pg^{ll}$, where $t+p+tp\geq 0,p+1\geq 0$ and $%
k<l$. The operator $g(t)$ is positive hence $F(g(t))$ is also positive.
Restrict $F(g(t))$ to the subspace $\langle e_{k}\otimes f_{i},e_{l}\otimes
f_{i}\rangle $. In this basis $F(g(t))$ has the form 
\[
\left( 
\begin{array}{cc}
F(g(t))_{ii}^{kk} & F(g(t))_{ii}^{lk} \\ 
F(g(t))_{ii}^{kl} & F(g(t))_{ii}^{ll}
\end{array}
\right) =\left( 
\begin{array}{cc}
F(g^{kl})_{ii}^{kk}+tF(g^{kk})_{ii}^{kk} & F(g^{kl})_{ii}^{lk} \\ 
F(g^{kl})_{ii}^{kl} & F(g^{kl})_{ii}^{ll}+pF(g^{ll})_{ii}^{ll}
\end{array}
\right) . 
\]
This matrix is positive, hence 
\[
(F(g^{kl})_{ii}^{kk}+tF(g^{kk})_{ii}^{kk})(F(g^{kl})_{ii}^{ll}+pF(g^{ll})_{ii}^{ll})\geq (F(g^{kl})_{ii}^{kl})^{2} 
\]
(note that $F(g^{kl})_{ii}^{kl}$, $F(g^{kl})_{ii}^{kk}+tF(g^{kk})_{ii}^{kk}$%
, and $F(g^{kl})_{ii}^{ll}+pF(g^{ll})_{ii}^{ll}$ are real and non-negative).
We apply Lemma \ref{pq} with $a=F(g^{kk})_{ii}^{kk}$, $b=F(g^{kl})_{ii}^{kl}$%
, $c=F(g^{ll})_{ii}^{ll}$, which gives us 
\[
F(g^{kk})_{ii}^{kk}=F(g^{ll})_{ii}^{ll}\leq F(g^{kl})_{ii}^{kl}. 
\]
Taking into account the fact that $\sum_{i=1}^{\infty
}F(g^{kk})_{ii}^{kk}=\sum_{i=1}^{\infty
}F(g^{ll})_{ii}^{ll}=\sum_{i=1}^{\infty }F(g^{kl})_{ii}^{kl}=1$ we get 
\[
F(g^{kk})_{ii}^{kk}=F(g^{ll})_{ii}^{ll}=F(g^{kl})_{ii}^{kl}. 
\]

Secondly, let us show that the remaining elements of the non-zero components
of $F(g^{kl}),k\leq l$ are equal. For that purpose we prove 
\[
F(g^{kk})_{ij}^{kk}=F(g^{ll})_{ij}^{ll}=F(g^{kl})_{ij}^{kl}\;\mbox{ for all }%
\,i\neq j\mbox{ and all }\,k<l 
\]
using again the operator $g(t)$. Denote $a_{i}=F(g^{kl})_{ii}^{kk}$.
Restrict $F(g(t))$ to the subspace $\langle e_{k}\otimes f_{i},e_{k}\otimes
f_{j},e_{l}\otimes f_{j}\rangle $. In this basis $F(g(t))$ has the form 
\begin{eqnarray*}
&&\left( 
\begin{array}{ccc}
F(g(t))_{ii}^{kk} & F(g(t))_{ji}^{kk} & F(g(t))_{ji}^{lk} \\ 
F(g(t))_{ij}^{kk} & F(g(t))_{jj}^{kk} & F(g(t))_{jj}^{lk} \\ 
F(g(t))_{ij}^{kl} & F(g(t))_{jj}^{kl} & F(g(t))_{jj}^{ll}
\end{array}
\right) \\
&=&\left( 
\begin{array}{ccc}
F(g^{kl})_{ii}^{kk}+tF(g^{kk})_{ii}^{kk} & 
F(g^{kl})_{ji}^{kk}+tF(g^{kk})_{ji}^{kk} & F(g^{kl})_{ji}^{lk} \\ 
F(g^{kl})_{ij}^{kk}+tF(g^{kk})_{ij}^{kk} & 
F(g^{kl})_{jj}^{kk}+tF(g^{kk})_{jj}^{kk} & F(g^{kl})_{jj}^{lk} \\ 
F(g^{kl})_{ij}^{kl} & F(g^{kl})_{jj}^{kl} & 
F(g^{kl})_{jj}^{ll}+pF(g^{ll})_{jj}^{ll}
\end{array}
\right) \\
&=&\left( 
\begin{array}{ccc}
a_{i}+ta_{i} & x+ty & x \\ 
\bar{x}+t\bar{y} & a_{j}+ta_{j} & a_{j} \\ 
\bar{x} & a_{j} & a_{j}+pa_{j}
\end{array}
\right) =A.
\end{eqnarray*}
If $a_{j}=0$ then obviously $x=y=0$ as A is positive. If $a_{j}\neq 0$ then 
\begin{eqnarray*}
\mbox{ det}A &=&a_{i}a_{j}^{2}(1+t)^{2}(1+p)+(x+ty)a_{j}\bar{x}+x(\bar{x}+t%
\bar{y})a_{j}-xa_{j}(1+t)\bar{x} \\
&&-(x+ty)(\bar{x}+t\bar{y})a_{j}(1+p)-a_{i}a_{j}^{2}(1+t) \\
&=&-\frac{a_{j}}{1+t}|x(1+t)-(x+ty)|^{2}\geq 0\Longrightarrow x=y.
\end{eqnarray*}
This means that 
\[
F(g^{kl})_{ij}^{kk}=F(g^{kk})_{ij}^{kk}. 
\]
\bigskip

\noindent \textbf{Step 6.} Firstly, let us show that the main diagonals of
non-zero components of all $F(g^{kl\ast }),k<l$, satisfy the equality 
\[
F(g^{kk})_{ii}^{kk}=-iF(g^{kl\ast })_{ii}^{kl}=iF(g^{kl\ast
})_{ii}^{lk}=F(g^{ll})_{ii}^{ll}. 
\]
Thereby we use the same arguments as in the previous step considering the
operator $g^{\ast }(t)=g^{kl\ast }+tg^{kk}+pg^{ll}$, where $t+p+tp\geq
0,p+1\geq 0$ and $k<l$. The operator $g^{\ast }(t)$ is positive, hence $%
F(g^{\ast }(t))$ is also positive. Restrict $F(g^{\ast }(t))$ to the
subspace $\langle e_{k}\otimes f_{i},e_{l}\otimes f_{i}\rangle $. In this
basis $F(g^{\ast }(t))$ has the form 
\begin{equation}
\left( 
\begin{array}{cc}
F(g^{\ast }(t))_{ii}^{kk} & F(g^{\ast }(t))_{ii}^{lk} \\ 
F(g^{\ast }(t))_{ii}^{kl} & F(g^{\ast }(t))_{ii}^{ll}
\end{array}
\right) =\left( 
\begin{array}{cc}
F(g^{kl\ast })_{ii}^{kk}+tF(g^{kk})_{ii}^{kk} & -F(g^{kl\ast })_{ii}^{kl} \\ 
F(g^{kl\ast })_{ii}^{kl} & F(g^{kl\ast })_{ii}^{ll}+pF(g^{ll})_{ii}^{ll}
\end{array}
\right) .  \label{m.6}
\end{equation}
Note that $-iF(g^{kl\ast })_{ii}^{kl}$, $F(g^{kl\ast
})_{ii}^{kk}+tF(g^{kk})_{ii}^{kk}$, and $F(g^{kl\ast
})_{ii}^{ll}+pF(g^{ll})_{ii}^{ll}$ are real and non-negative. The matrix (%
\ref{m.6}) is positive, hence 
\[
(F(g^{kl\ast })_{ii}^{kk}+tF(g^{kk})_{ii}^{kk})(F(g^{kl\ast
})_{ii}^{kk}+pF(g^{ll})_{ii}^{ll})\geq -|F(g^{kl\ast
})_{ii}^{kl}|^{2}=|-iF(g^{kl\ast })_{ii}^{kl}|^{2}. 
\]
We apply Lemma \ref{pq} with $a=F(g^{kk})_{ii}^{kk}$, $b=-iF(g^{kl\ast
})_{ii}^{kl}$, $c=F(g^{ll})_{ii}^{ll}$, which gives us 
\[
F(g^{kk})_{ii}^{kk}=F(g^{ll})_{ii}^{ll}\leq -iF(g^{kl\ast })_{ii}^{kl}. 
\]
Taking into account the fact that $\sum_{i=1}^{\infty
}F(g^{kk})_{ii}^{kk})=\sum_{i=1}^{\infty
}F(g^{ll})_{ii}^{ll}=-\sum_{i=1}^{\infty }iF(g^{kl\ast })_{ii}^{kl}=1$ we
get 
\[
F(g^{kk})_{ii}^{kk}=-iF(g^{kl\ast })_{ii}^{kl}=iF(g^{kl\ast
})_{ii}^{lk}=F(g^{ll})_{ii}^{ll}\;\mbox{ for all }k<l. 
\]

Secondly, let us show that the remaining elements of the non-zero components
of $F(g^{kl\ast }),\,k<l$, satisfy 
\[
F(g^{kk})_{ij}^{kk}=F(g^{kl\ast })_{ij}^{kk}\quad \mbox{\rm  if }\,i\neq j 
\]
using again the operator $g^{\ast }(t)$. Denote $a_{i}=F(g^{kl\ast
})_{ii}^{kk}$. Restrict $F(g^{\ast }(t))$ to the subspace $\langle
e_{k}\otimes f_{i},e_{k}\otimes f_{j},e_{l}\otimes f_{j}\rangle $. In this
basis $F(g^{\ast }(t))$ has the form 
\begin{eqnarray*}
&&\left( 
\begin{array}{ccc}
F(g^{\ast }(t))_{ii}^{kk} & F(g^{\ast }(t))_{ji}^{kk} & F(g^{\ast
}(t))_{ji}^{lk} \\ 
F(g^{\ast }(t))_{ij}^{kk} & F(g^{\ast }(t))_{jj}^{kk} & F(g^{\ast
}(t))_{jj}^{lk} \\ 
F(g^{\ast }(t))_{ij}^{kl} & F(g^{\ast }(t))_{jj}^{kl} & F(g^{\ast
}(t))_{jj}^{ll}
\end{array}
\right) \\
&=&\left( 
\begin{array}{ccc}
F(g^{kl\ast })_{ii}^{kk}+tF(g^{kk})_{ii}^{kk} & F(g^{kl\ast
})_{ji}^{kk}+tF(g^{kk})_{ji}^{kk} & F(g^{kl\ast })_{ji}^{lk} \\ 
F(g^{kl\ast })_{ij}^{kk}+tF(g^{kk})_{ij}^{kk} & F(g^{kl\ast
})_{jj}^{kk}+tF(g^{kk})_{jj}^{kk} & F(g^{kl\ast })_{jj}^{lk} \\ 
F(g^{kl\ast })_{ij}^{kl} & F(g^{kl\ast })_{jj}^{kl} & F(g^{kl\ast
})_{jj}^{ll}+pF(g^{ll})_{jj}^{ll}
\end{array}
\right) \\
&=&\left( 
\begin{array}{ccc}
a_{i}+ta_{i} & x+ty & -ix \\ 
\bar{x}+t\bar{y} & a_{j}+ta_{j} & -ia_{j} \\ 
i\bar{x} & ia_{j} & a_{j}+pa_{j}
\end{array}
\right) =A.
\end{eqnarray*}
If $a_{j}=0$ then obviously $x=y=0$ as A is positive. If $a_{j}\neq 0$ then,
analogously to the previous step, 
\[
\mbox{ det}A=-\frac{a_{j}}{1+t}|x(1+t)-(x+ty)|^{2}\geq 0\Longrightarrow x=y. 
\]
This means that 
\[
F(g^{kl\ast })_{ij}^{kk}=F(g^{kk})_{ij}^{kk}. 
\]
\bigskip

Denote $a_{ij}=F(g^{11})_{ij}^{11}$ and consider $\rho _{E}\in \mathcal{L}%
_{1}^{+}(H_{S})$ that has the form $a_{ij}$ in the basis $\{e_{i},i\in \mathbb{N%
}\}$. It is easy to see now that $F(\rho )=\rho \otimes \rho _{E}$ for each $%
\rho \in \mathcal{L}_{1}(H)$. The theorem is proved.\hfill $\Box $ \smallskip

\begin{remark}
The theorem implies that the linear lifting $F$ is continuous.
\end{remark}

\begin{remark}
If we skip the constraint $\mathrm{tr}_{H_{E}}F(\rho )=\rho $, more general
liftings are possible. Let $\rho _{E}\in \mathcal{D}(H_{E})$ be a reference
state, and $K_{n}$ a family of bounded operators in $H$ which satisfy $%
\sum_{n}K_{n}^{+}K_{n}=Id$, then 
\begin{equation}
\rho \longmapsto F(\rho )=\sum_{n}K_{n}\left( \rho \otimes \rho _{E}\right)
K_{n}^{+}
\end{equation}
is a linear and continuous mapping $\mathcal{D}(H_{S})\rightarrow \mathcal{D}%
(H).$ Such liftings are used in general investigations of the process of
measurement \cite{Kraus:1983} and in information theory, see e. g. \cite
{Accardi/Ohya:1999}.
\end{remark}

\begin{remark}
It is well known that any mixed state $\rho $ of a system $\mathcal{S}$ can
be obtained as the reduced state of a pure state in an extended system $%
\mathcal{S}\times \mathcal{E}$, if only $\dim H_{E}\geq \dim H_{S}$, see
e.g. \cite{Peres:1993}. But due to Theorem \ref{lifting} the pure state
cannot depend linearly on the state $\rho $. The representation by a pure
state is actually a generalization of the classical Gram's theorem from
linear algebra. To see this let $H_{S}$ be realized as $\mathcal{L}%
_{2}(\Omega ,\mathcal{B}_{\Omega },\mu _{\Omega })$ where $\Omega $ is a set
, $\mathcal{B}_{\Omega }$ is a $\sigma $-algebra of its subsets, $\mu
_{\Omega }$ a non-negative $\sigma $-additive measure on $\mathcal{B}%
_{\Omega }$. Then the space $H=H_{S}\otimes H_{E}$ is isomorphic to the
space $\mathcal{L}_{2}(\Omega ,\mathcal{B}_{\Omega },\mu _{\Omega },H_{E})$
of $H_{E}$-valued Bochner square $\mu _{\Omega }$-integrable functions on $%
\Omega $. The corresponding isomorphic map $H_{S}\otimes H_{E}\rightarrow 
\mathcal{L}_{2}(\Omega ,\mathcal{B}_{\Omega },\mu _{\Omega },H_{E})$ is
denoted by $\varphi $. On the other hand, the space $H_{S}\otimes H_{S}$ can
be realized as $\mathcal{L}_{2}(\Omega \times \Omega ,\mathcal{B}_{\Omega
}\otimes \mathcal{B}_{\Omega },\mu _{\Omega }\otimes \mu _{\Omega })$, and
hence the space $\mathcal{L}_{1}^{+}(H_{S})$ can be considered as a vector
subspace of the latter space which includes all Hilbert-Schmidt operators in 
$H_{S}$. Any normalized vector $a\in H_{S}\otimes H_{E},\,\left\| a\right\|
=1$, spans a one-dimensional subspace of $H_{S}\otimes H_{E}$ and defines a
unique projection operator $P_{a}\in \mathcal{D}(H_{S}\otimes H_{E})$. If $%
f_{a}\in \mathcal{L}_{2}(\Omega ,\mathcal{B}_{\Omega },\mu _{\Omega },H_{E})$
is defined by $f_{a}=\varphi (a)$ then the reduced state of pure state $%
P_{a} $ is given by 
\begin{equation}
S(\omega _{1},\omega _{2})=\left\langle f_{a}(\omega _{1}),f_{a}(\omega
_{2})\right\rangle _{H_{E}}.  \label{gram}
\end{equation}
Now the generalization of Gram's theorem can be formulated as follows: For
any $S\in $ $\mathcal{L}_{1}^{+}(H_{S})$ there exists a vector $a\in
H_{S}\otimes H_{E},\,\left\| a\right\| =1$, for which (\ref{gram}) holds. If 
$\Omega $ is a finite set and $\mu _{\Omega }$ is the counting measure, we
obtain the classical Gram's theorem.
\end{remark}

\section{\textbf{REDUCING OBSERVABLES}}

The problem of linear liftings of states is closely related to the problem
of reducing observables of the total system $H$ to observables of the
subsystem $H_{S}$.

\begin{lemma}
Let $F:\mathcal{L}_{1}(H_{S})\longrightarrow \mathcal{L}_{1}(H_{S}\otimes
H_{E})$ be a continuous mapping and let $F^{\ast }:\mathcal{L}(H_{S}\otimes
H_{E})\longrightarrow \mathcal{L}(H_{S})$ be its adjoint mapping; then $%
F^{\ast }(B\otimes Id_{E})=B$ for all $B\in \mathcal{L}(H_{S})$ iff $\mathrm{%
tr}_{H_{E}}F(\rho )=\rho $ for all ${\rho \in \mathcal{L}_{1}(H_{S})}$.
\end{lemma}

\noindent \textbf{Proof} \ If $B\in \mathcal{L}(H_{S})$ then, according to
the definition of the duality between $\mathcal{L}(H)$ and $\mathcal{L}%
_{1}(H)$, $\langle B\otimes Id_{E},F(\rho )\rangle =tr_{H}(B\otimes
Id_{E})F(\rho )=\mathrm{tr}_{H_{S}}B\rho =\langle B,\rho \rangle $. This
identity together with the definition of the duality between $\mathcal{L}%
(H_{S})$ and $\mathcal{L}_{1}(H_{S})$ implies that 
\begin{equation}
F^{\ast }(B\otimes Id)=B.  \label{star}
\end{equation}
On the other hand, if $F^{\ast }$ satisfies (\ref{star}) then, for $B\in 
\mathcal{L}(H_{S})$ and $\rho \in \mathcal{L}_{1}(H_{S})$, 
\begin{equation}
\langle B\otimes Id_{E},F(\rho )\rangle =\langle F^{\ast }(B\otimes
Id_{E}),\rho \rangle =\langle B,\rho \rangle =\mathrm{tr}_{H_{S}}B\rho .
\end{equation}

But $\langle B\otimes Id_{E},F(\rho )\rangle =\mathrm{tr}_{H_{S}}B(\mathrm{tr%
}_{H_{E}}F(\rho ))$. Hence $\langle B,\rho \rangle =\mathrm{tr}_{H_{S}}B\rho
=\langle B,\mathrm{tr}_{H_{E}}F(\rho )\rangle $, and as the latter identity
holds for any $B$, we finally obtain $\rho =\mathrm{tr}_{H_{E}}F(\rho )$.
The lemma is proved. \hfill $\Box$ \smallskip

Theorem 1 and Lemma 2 imply the following theorem.

\begin{theorem}
If $R:\mathcal{L}(H_{S}\otimes H_{E})\rightarrow \mathcal{L}(H_{S})$ is a
linear mapping, continuous in the ultraweak or $\left( \sigma (\mathcal{L}%
(H),\mathcal{L}_{1}(H)),\sigma (\mathcal{L}(H_{S}),\mathcal{L}%
_{1}(H_{S}))\right) $ topology, see e.g. Sec. VI.6 of \cite{Reed/Simon:1972}%
, and if \newline
$R(B\otimes Id_{E})=B$ is true for all $B\in \mathcal{L}(H_{S})$ then there
exists an element $\rho _{E}\in \mathcal{D}(H_{S})$ such that $R(A)=\mathrm{%
tr}_{H_{E}}A(Id_{S}\otimes \rho _{E})$ for all $A\in \mathcal{L}(H)$.
\end{theorem}

\section{\textbf{PROBABILITY MEASURES}}

The classical analog of the case considered in Theorem 1 is much simpler and
admits non-factorizing answers. Let $T$ be a topological space, then $%
\mathcal{C}_{b}(T)$ is the vector space of all bounded continuous functions
on $T$, $\mathcal{M}(T)$ is the vector space of all Borel (signed) measures
on $T$ equipped with the topology $\sigma (\mathcal{M}(T),\mathcal{C}%
_{b}(T)) $, and $\mathcal{M}_{p}(T)$ is the closed convex set of probability
measures on $T$. The Dirac measure at point $t\in T$ will be denoted by $%
\delta _{t}$. Let $Q$ and $P$ be topological spaces, $E=Q\times P$ the
product space, and $\mathcal{G}:\mathcal{M}_{p}(E)\rightarrow \mathcal{M}%
_{p}(Q)$ be the mapping induced by the projection $\mathrm{pr}%
_{Q}:E\rightarrow Q$. The mapping $\mathcal{G}$ can be (uniquely) extended
by linearity to an $\mathbb{R}$-linear mapping $\mathcal{M}(E)\rightarrow 
\mathcal{M}(Q)$. For any measure $\mu \in \mathcal{M}(E)$ the measure $%
\mathcal{G}\mu \in \mathcal{M}(Q)$ is called the marginal of $\mu $. The
right inverse of $\mathcal{G}$ will be called a lifting.

\begin{lemma}
\label{classical}Let $f:Q\rightarrow \mathcal{M}_{p}(E)$ be a continuous
function such that $\mathcal{G}f(q)=\delta _{q}$, then the mapping $\mathcal{%
F}:\mathcal{M}(Q)\rightarrow \mathcal{M}(E)$ defined by 
\begin{equation}
\mathcal{F}\upsilon :=\int_{Q}f(q)\upsilon (dq)  \label{ch.1}
\end{equation}
is a linear lifting. Any linear lifting has this representation.
\end{lemma}

\noindent \textbf{Proof} \ Take the Dirac measure $\delta _{q}$ then the
integral is $\mathcal{F}\delta _{q}=f(q)\in \mathcal{M}_{p}(E)$ and we have $%
\mathcal{GF}\delta _{q}=\mathcal{G}f(q)=\delta _{q}$. The general case
follows by linearity and continuity. On the other hand, if $G$ is a linear
lifting, then (\ref{ch.1}) follows with the function $f(q)=\mathcal{F}\delta
_{q}$.\hfill $\Box $ \smallskip 

If $f(q)$ factorizes into $f(q)=\delta _{q}\times \chi $ with $\chi \in 
\mathcal{M}(P)$, the lifting (\ref{ch.1}) factorizes into $\mathcal{F}%
(\upsilon )=\upsilon \times \chi $. But one can obviously choose
non-factorizing functions $f(q)$ such that $\mathcal{F}(\upsilon )$ is not a
product measure. To give an explicit example we split $Q$ into two disjoint
measurable sets $Q=Q_{1}\cup Q_{2}$ and denote by $\chi _{1}(q)$ and $\chi
_{2}(q)$ the characteristic functions of the sets $Q_{1}$ and $Q_{2}$. Then 
\begin{equation}
f(q)=\chi _{1}(q)\delta _{q}\times \delta _{p_{1}}+\chi _{2}(q)\delta
_{q}\times \delta _{p_{2}}  \label{ch.2}
\end{equation}
with two points $p_{j}\in P,\,j=1,2,\;p_{1}\neq p_{2}$, yields an example of
a non-factorizing lifting.

The state space $\mathcal{D}(H)$ of a quantum mechanical system is a closed
convex set with the pure states $\mathcal{P}(H)$ as extremal points. Any $%
W\in \mathcal{D}(H)$ can be represented by the Choquet integral \cite
{Choquet:1969} 
\begin{equation}
W=\int_{\mathcal{P}(H)}P\,\mu (dP)  \label{ch.3}
\end{equation}
where $\mu (dP)$ is a -- in general non-unique -- measure in the convex set $%
\mathcal{M}_{p}(\mathcal{P}(H))$ of probability measures on $\mathcal{P}(H)$%
, see e. g. \cite{Kupsch:1998}. This representation relates the quantum
mechanical state space with the space of probability measures, and one might
ask whether it is possible to find an affine linear mapping $\gamma :%
\mathcal{D}(H)\rightarrow \mathcal{M}(\mathcal{P}(H))$ such that (\ref{ch.1}%
) is valid for all $W\in \mathcal{D}(H)$ with the measure $\mu (dP)=\gamma
_{W}(dP)$.

\begin{theorem}
\label{simplex}There does not exist an affine linear mapping $\gamma :%
\mathcal{D}(H)\rightarrow \mathcal{M}_{p}(\mathcal{P}(H))$ such that the
representation (\ref{ch.1}) holds for all $W\in \mathcal{D}(H)$ with $\mu
(dP)=(\gamma W)(dP)$.
\end{theorem}

\noindent \textbf{Proof} \ If such a mapping $\gamma $ exists then any pure
state has to be represented by an atomic measure on the one-point set
containing just this pure state. Moreover this mapping can be extended to an 
$\mathbb{R}$-linear mapping $\gamma :\mathcal{L}_{1}^{a}(H)\rightarrow \mathcal{%
M}(\mathcal{P}(H))$. Since there are finite sets of pure states which are
linearly dependent in $\mathcal{L}_{1}^{a}(H)$ -- e.g. any four projection
operators on the Hilbert subspace $\mathbb{C}^{2}$ of $\mathcal{H}$ -- whereas
the set of atomic measures is linear independent in $\mathcal{M}(\mathcal{P}%
(H))$ we obtain contradiction to the linearity of $\gamma $.\hfill $\Box $
\smallskip

The proof given here exploits the different structures of the convex sets $%
\mathcal{D}(H)$ and \newline
$\mathcal{M}_{p}(\mathcal{P}(H))$: the space of measures is a simplex
whereas $\mathcal{D}(H)$ not. Theorem \ref{simplex} is also closely related
to our main Theorem 1, it is actually a consequence of it. To see that
implication assume such an affine linear mapping $\gamma $ exists. Then the
lifting problem of Sec. 2. has the following solution in contradiction to
Theorem 1.

In the first step the statistical operator $\rho \in \mathcal{D}(H_{S})$ is
mapped onto the measure $\gamma \rho \in \mathcal{M}_{p}(\mathcal{P}(H_{S}))$%
. Following Lemma \ref{classical} we can lift this measure to a measure $%
\sigma \in \mathcal{M}_{p}(\mathcal{P}(H_{S})\times \mathcal{P}(H_{E}))$.
Thereby we can choose a lifting such that $\sigma $ is not a product
measure, take e.g. (\ref{ch.2}). The operator 
\begin{equation}
W=\int_{\mathcal{P}(H_{S})\times \mathcal{P}(H_{E})}P_{S}\otimes
P_{E}\,\sigma (dP_{S}\times dP_{E})  \label{ch.4}
\end{equation}
has the partial trace $\mathrm{tr}_{H_{E}}W=\int_{\mathcal{P}%
(H_{S})}P_{S}\,(\gamma \rho )(dP_{S})=\rho $. All steps of the mapping $\rho
\rightarrow W$ are affine linear. Since the measure $\sigma $ does not
factorize, the statistical operator $W$ has not the product form $\rho
\otimes \rho _{E}$, and we have obtained a contradiction to Theorem 1.

In addition to the representation of states by a probability distribution on
the set of pure states there exists a representation of any state by a
random vector distributed by a probability measure on the Hilbert space.
Such a representation is used in the theory of Schr\"{o}dinger (-Belavkin)
stochastic equations (see \cite{AKS:1997}, \cite{Belavkin/Smol:1998} and
references therein), which gives both, a phenomenological description of
continuous measurements, and a Markovian approximations for the reduced
dynamics.

By $\mathcal{M}(H)$ we denote the space of all $\sigma $-additive signed
measures on the $\sigma $-algebra of Borel subsets of $H$. The space of
probability measures on $H$ is denoted by $\mathcal{M}_{p}(H)$, the set of
all measures concentrated on $H\backslash \{0\}$ by $\mathcal{M}^{0}(H)$,
and the set of all probability measures concentrated on $H\backslash \{0\}$
by $\mathcal{M}_{p}^{0}(H)=\mathcal{M}^{0}(H)\cap \mathcal{M}_{p}(H)$.

In the theory of stochastic Schr\"{o}dinger equations a probability measure $%
\nu \in \mathcal{M}_{p}^{0}(H)$ represents the state $B\in \mathcal{D}(H)$
if 
\begin{equation}
\int\limits_{H}\langle z,Az\rangle \Vert z\Vert ^{-2}\nu (dz)=\omega
_{B}(A)\equiv \mathrm{tr}_{H}AB  \label{sch.1}
\end{equation}
is valid for all observables $A\in \mathcal{L}(H)$. Thereby any measure $\nu
\in \mathcal{M}_{p}^{0}(H)$ represents a state, and any state $W\in \mathcal{%
D}(H)$ can be represented by such a measure.

For the proof of the first statement take $A\in \mathcal{L}(H)$. Then the
function $\left| \left\langle z,Az\right\rangle \right| \Vert z\Vert ^{-2}$
is bounded by $\Vert A\Vert $ for all $z\neq 0$, and the integral $\omega
_{\nu }(A):=\int\limits_{H}\Vert z\Vert ^{-2}\langle z,Az\rangle \nu (dz)$
is defined. Moreover, it is easy to see that all demands of Gleason's
theorem, see Remark \ref{Gleason}, are fulfilled. Hence there exists a state 
$W\in \mathcal{D}(H)$ such that $\omega _{\nu }(A)=\mathrm{tr}_{H}AW$.

On the other hand, given a statistical operator a probability measure for
the representation (\ref{sch.1}) can be constructed as follows. For any $%
B\in \mathcal{D}(H)$, let $\nu _{B}^{0}\in \mathcal{M}_{p}^{0}(H)$ be a
probability measure with the correlation operator $B$, i.e. for all $%
z_{1},z_{1}\in H$ the identity $\langle z_{1},Bz_{2}\rangle =\int \langle
z_{1},z\rangle \langle z,z_{2}\rangle \nu _{B}^{0}(dz)$ is true. It is worth
noticing that among the measures $\nu _{B}^{0}$ there exist precisely one
Gaussian measure with zero mean value. The positive measure $\nu _{B}\in 
\mathcal{M}^{0}(H)$ is then defined by $\nu _{B}=\langle z,z\rangle \nu
_{B}^{0}=\left\| z\right\| ^{2}\nu _{B}^{0}$; i.e. for any Borel subset $%
\mathcal{A}$ of $H^{\mathbb{R}}$ we have $\nu _{B}(\mathcal{A})=\int\limits_{%
\mathcal{A}}\langle z,z\rangle \nu _{B}^{0}(dz)$. The identity $\mathrm{tr}%
_{H}\,B=1$ implies that $\nu _{B}$ is a probability measure on $H$; in fact $%
\nu _{B}(H)=\int \langle z,z\rangle \nu _{B}^{0}(dz)=\mathrm{tr}_{H}\,B=1$.
For any observable $A\in \mathcal{L}(H)$ the function $\,H\longmapsto \mathbb{R}%
^{1}:\;z\longmapsto \frac{1}{\Vert z\Vert ^{2}}\langle z,Az\rangle $ is a
random variable on the probability space $(H,\nu _{B})$. The mean value $%
\bar{A}$ of this random variable 
\[
\bar{A}=\int\limits_{H}\langle z,Az\rangle \Vert z\Vert ^{-2}\nu
_{B}(dz)=\int\limits_{H}\langle z,Az\rangle \nu _{B}^{0}(dz)=\mathrm{tr}%
_{H}\,AB 
\]
is exactly the expectation of the observable $A$ in the state $B\in \mathcal{%
D}(H)$. Hence the measure $\nu _{B}\in \mathcal{M}_{p}^{0}(H)$ represents
the state $B$.

There exists an affine linear mapping from the measures $\upsilon \in 
\mathcal{M}_{p}^{0}(H)$ into the set of measures of the Choquet
representation. Let $\varphi :H\backslash \{0\}\rightarrow \mathcal{P}(H)$
be the mapping $a\mapsto P_{a}$, where $P_{a}$ is the projection operator
onto the subspace $\left\{ \lambda a\mid \lambda \in \mathbb{C}\right\} $, i.e. 
$P_{a}b=\left\langle b\mid a\right\rangle \left\| a\right\| ^{-2}a$ for all $%
b\in H$. Then the measure $\nu \varphi ^{-1}\in \mathcal{M}_{p}(\mathcal{P}%
(H))$ is defined by $\nu \varphi ^{-1}(\mathcal{R})=\nu \left( \varphi ^{-1}(%
\mathcal{R})\right) $ for any measurable set $\mathcal{R}\subset \mathcal{P}%
(H)$ of projection operators. This mapping $\nu \mapsto \nu \varphi ^{-1}$
is affine linear. If $\nu \in \mathcal{M}_{p}^{0}$ represents a state $W\in 
\mathcal{D}(H)$, then (\ref{sch.1}) and the definition of $\nu \varphi ^{-1}$
yield 
\[
\left\langle \,z_{1}\mid Wz_{2}\right\rangle \stackrel{(\ref{sch.1})}{=}%
\int_{H}\left\langle \,z_{1}\mid z\right\rangle \left\langle \,z\mid
z_{2}\right\rangle \Vert z\Vert ^{-2}\nu (dz)=\int_{\mathcal{P}%
(H)}\left\langle \,z_{1}\mid Pz_{2}\right\rangle (\nu \varphi ^{-1})(dP). 
\]
But that means $W=\int_{\mathcal{P}(H)}P\,(\nu \varphi ^{-1})(dP)$, and $\nu
\varphi ^{-1}$ is the Choquet measure of the state $W$.

The measures in the representation (\ref{sch.1}) are highly nonunique; the
arbitrariness is even larger than in the case of the Choquet representation,
and one might ask again for an affine linear lifting $\mathcal{D}%
(H)\rightarrow \mathcal{M}_{p}^{0}(H)$. But assume such an affine linear
lifting $\gamma :\,\mathcal{D}(H)\rightarrow \mathcal{M}_{p}^{0}(H)$ exists,
then it induces an affine linear lifting $\,\mathcal{D}(H)\rightarrow 
\mathcal{M}_{p}(\mathcal{P}(H))$ by $W\mapsto \gamma (W)\mapsto \left(
\gamma (W)\right) \varphi ^{-1}$ and we have obtained a contradiction to
Theorem \ref{simplex}.

\begin{corollary}
There does not exist an affine linear mapping $\gamma :\,\mathcal{D}%
(H)\rightarrow \mathcal{M}_{p}^{0}(H)$ such that for any $W\in \mathcal{D}%
(H) $ the measure $\gamma (W)$ represents the state $W$.
\end{corollary}

\medskip

\begin{center}
\textbf{Acknowledgment}
\end{center}

\noindent This work was done during a stay of O. G. Smolyanov at the
University of Kaiserslautern. OGS would like to thank the Deutsche
Forschungsgemeinschaft (DFG) for financial support.


\end{document}